\documentclass{sigchi}

\usepackage{subfigure}
\usepackage{balance}  
\usepackage{graphics} 
\usepackage{times}    
\usepackage{url}      

\makeatletter
\def\url@leostyle{%
  \@ifundefined{selectfont}{\def\UrlFont{\sf}}{\def\UrlFont{\small\bf\ttfamily}}}
\makeatother
\newcommand{\Fig}[1] {Figure~\ref{fig:#1}}

\renewcommand \paragraph[1] {\vspace{0.05cm} \textbf{#1}}

\def\pprw{8.5in}
\def\pprh{11in}

\usepackage[pdftex]{hyperref}
\hypersetup{
pdftitle={Camera Swarm: Collaborative and Synchronized Multi-Phone Photography},
pdfauthor={LaTeX},
pdfkeywords={SIGCHI, proceedings, archival format},
bookmarksnumbered,
pdfstartview={FitH},
colorlinks,
citecolor=black,
filecolor=black,
linkcolor=black,
urlcolor=black,
breaklinks=true,
}

\begin{document}

\title{PanoSwarm: Collaborative and Synchronized \\ Multi-Device Panoramic Photography}

\numberofauthors{3}
\author{
  \alignauthor Yan Wang \\
    \affaddr{Columbia University}\\
    \affaddr{2960 Broadway, New York, NY, USA}\\
    \email{yanwang@ee.columbia.edu}\\
  \alignauthor Sunghyun Cho \\
    \affaddr{Samsung Electronics}\\
    \affaddr{95, Samsung 2-ro, Giheung-gu, Yongin-si, Gyeonggi-do, Korea, 446-811}\\
    \email{sodomau@postech.ac.kr}\\
  \alignauthor Jue Wang \\
    \affaddr{Adobe Systems}\\
    \affaddr{801 N 34th Street, Seattle, WA, USA}\\
    \email{juewang@adobe.com}\\
  \alignauthor Shih-Fu Chang\\
    \affaddr{Columbia University}\\
    \affaddr{2960 Broadway, New York, NY, USA}\\
    \email{sfchang@ee.columbia.edu}\\
}

\maketitle

\begin{abstract}
Taking a picture has been traditionally a one-person's task. 
In this paper we present a novel system that allows multiple mobile devices to work collaboratively in a synchronized fashion to capture a panorama of a highly dynamic scene, creating an entirely new photography experience that encourages social interactions and teamwork. 
Our system contains two components: a client app that runs on all participating devices, and a server program that monitors and communicates with each device. 
In a capturing session, the server collects in realtime the viewfinder images of all devices and stitches them on-the-fly to create a panorama preview, which is then streamed to all devices as visual guidance. 
The system also allows one camera to be the host and to send direct visual instructions to others to guide camera adjustment. 
When ready, all devices take pictures at the same time for panorama stitching. Our preliminary study suggests that the proposed system can help users capture high quality panoramas with an enjoyable teamwork experience. 
\end{abstract}



\section{Introduction}\label{sec:intro}

Photography has been largely a one person's task since the invention of the camera in the 18th century.
While setting up the scene may require group effort, the camera has always been designed for a single photographer who controls all the key factors in the capturing process, from scene composition and camera setting to the shutter release time.
Thus, despite the rapid advances on camera hardware in recent years, the basic workflow of taking a picture, i.e. a single camera controlled by one photographer, largely remains unchanged. 

\begin{figure}[t]
    \centering
    \includegraphics[width=0.45\textwidth]{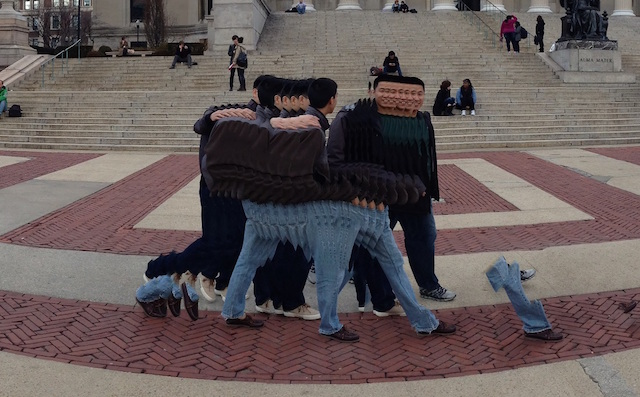}
    \caption{Severe ghosting artifacts observed in the conventional panorama stitching approaches.}
    \label{fig:failedpano}
    \vspace{-0.3cm}
\end{figure}

Unfortunately, a single camera cannot satisfy all the creative needs that consumers may have.
As a representative example, taking multiple shots and stitching them into a panorama has become a standard feature in most modern image capturing apps on mobile devices.
However, it does not work well for highly dynamic scenes that contain fast moving objects such as pedestrians or cars, which can easily lead to severe ghosting artifacts due to their fast motion in the capturing process (see \Fig{failedpano}).
To avoid such artifacts, it is desirable to have all photos covering different parts of the scene to be taken at exactly the same time, which is beyond the capability of a single consumer camera.  

In this paper, we propose a novel system that allows multiple regular mobile devices to dynamically form a team, and work collaboratively in a synchronized fashion to capture a high quality panorama of a dynamic scene.
Our system includes a networked server that continuously collects low-resolution viewfinder images from all participating devices during a capturing session, analyzes their content to discover the relative camera positions, and stitches them into a preview panorama.
On the client side, each camera continuously streams all the information from the server and presents it on a unique capturing interface.
On the interface each user can see not only how his/her own image contributes to the final result, but also a direct visualization of the scene coverage of other cameras.
Guided by the visualization, the user's task is to move the camera relative to others to increase the scene coverage of the panorama while avoiding gaps and holes in-between cameras. 
Our system also allows one user to become the ``host'', who can send visual guidance to other users on how to move their cameras. 
When all the cameras are ready, the host sends a signal through the server to capture all the final images at approximately the same time, and renders a high quality panorama from them as the final output.  
With the simultaneous photo capturing design, the notorious ghost effects in conventional single camera settings can now be alleviated. 

By also allowing a group of users to work as a team to capture a single photo, our system introduces an entirely new social photo capturing experience. 
The recent development of mobile device cameras enables consumers to take high quality pictures anywhere at anytime, and further allows them to be shared on social networks right after being captured, transforming photography into an essential component in our daily social lives.
The creative community also started to explore new ways to allow a group of users to collaboratively develop a photo story using photos from different users~\cite{4am,someone}. 
However, all these social interactions happen {\em after} the photos have been taken. 
In contrast, our system enables teamwork and user interactions {\em before} a photo is taken, when the users are at the same physical location together. 

The main design goal of the system is twofold:
(1) on the technical side, we aim to provide a mechanism for consumers to capture high quality panoramas using regular mobile devices, which cannot be achieved otherwise with existing software;
and (2) on the user experience side, we aim to provide an enjoyable social photography experience for consumers with different levels of photography skills.
To evaluate whether our system achieves the above goals, we conducted a preliminary user study by recruiting user groups to use the system to capture panoramas of real scenes.
The result suggests that our system is capable of capturing high quality panoramas, and the process is enjoyable.
It also provides us insights on how to further improve the user experience, and guidance for designing future larger scale studies.   

\section{Related Work}

\subsection{Panorama Stitching}
Panorama stitching has been extensively studied in computer vision and graphics in the last decade~\cite{Brown:2007}, and the underlying technologies have been significantly advanced to allow efficient implementations on mobile phones~\cite{Xiong:2010}, sometimes with realtime preview~\cite{PanoRealtime}.
There are many mature software products for creating panoramas as post-processing, such as Realviz Stitcher 5 and Canon PhotoStitch.
It also becomes a standard feature in the photo capturing modules of mobile platforms, such as Apple iOS 7 and Windows Phone 8.
All these solutions require the user to take a series of images that cover the scene, while maintaining some amount of partial overlap among the images for the purpose of alignment~\cite{Szeliski:2006}.

A well-known difficulty for panorama stitching is moving objects~\footnote{Top Panorama Problem \#3: Ghosting: $\;\;\;\;\;\;\;\;$      
http://www.theperfectpanorama.com/articles/problem-3-ghosting.html.}, which often cause ghosting artifacts as they appear in different positions in different images.
Although various computational solutions have been proposed to reduce the artifacts~\cite{Davis:1998}, their effectiveness and application range is limited, and some of them require additional user input that is tedious to provide~\cite{Agarwala:2004}.
Another solution to this problem is to develop special hardware systems that hook up multiple cameras to take all shots at the same time, such as the Mrotator One-Shot Photographic head produced by Agno's.  
However, these devices are expensive and bulky, designed for high-end professional tasks rather than consumer applications.  

Various applications and interfaces have also been developed for panorama stitching, especially on mobile devices.
Looking At You~\cite{LookAtYou} is an interface using mobile sensors such as gyroscopes and face tracking to help users view large imagery on mobile devices.
Videos can also be integrated within panoramic contexts, providing a browsing experience across both the temporal and spatial dimensions~\cite{PanoVideo}.
However, none of the works addresses collaborative capturing or the ghosting artifacts in the panoramas.

\subsection{Multi-Camera Arrays}

Building multi-camera array systems using inexpensive sensors has been extensively explored.
These systems can be used for high performance imaging, such as capturing high speed, high resolution and high dynamic range videos~\cite{Wilburn:2004,Wilburn:2005}, or the light field~\cite{Venkataraman:2013}.
These systems however are not designed for multi-user collaboration. 
Some of them contain large-scale, fixed support structures that cannot be easily moved around, and they all require special controlling hardware for sensor synchronization.
Our approach, in contrast, uses existing hardware and software infrastructure (e.g. wireless communication) for camera synchronization, thus can be easily deployed to consumers. More importantly, user interaction plays a central role in our capturing process, but not in traditional camera arrays.    

\subsection{Multi-Agent Robot System}
Multi-Agent robot systems are also related to our target scenario of a distributed system aiming at a common task.
RoboCup Soccer competition~\cite{robocup} is an influential competition for teams of robots to collaborate with the teammates in a soccer game.
There exists previous work on coordinating a group of quadrocopters to maximize the surveillance coverage by doing collaborative sensing and motion planning~\cite{KumarIJRR, KumarIROS}.
However, there was no human social interactions in such robot systems and the idea of using synchronized capturing to avoid motion artifacts was not considered.

\subsection{Collaborative Multi-User Interface}
In addition to autonomous robot systems, multi-user interface design is also a rising topic, especially in the HCI community.
The ``It's Mine, Don't Touch'' project~\cite{DontTouch} builds a large multi-touch display in the city center of Helsinki, Finland for the users to do massively parallel interaction, teamwork, and games. 
Exploration has also been done about how to enable multiple users to do painting and puzzle games on multiple mobile devices~\cite{MultiUserVideo}.
The above research observes that it is possible to naturally inspire social interactions from the users' conflicts in using the interfaces.
Our work is inspired by this observation, but it also differs from these interfaces in that we make panorama stitching as the primary goal, providing users a means to obtain motion-artifact-free panoramas, which has its unique technical challenges and cannot be done with the interfaces mentioned above.

\subsection{Collaborations in Photography}

Collaboration has deep roots in the history of photography, as shown in a recent art project that reconsiders the story of photography from the perspective of collaboration~\cite{Collaboration}. This study created a gallery of roughly one hundred photography projects in history and showed how ``photographers co-labor with each other and with those they photograph''.   

Recently advances on Internet and mobile technologies allow photographers that are remote to each other in space and/or time to collaboratively work on the same photo journalism or visual art creation project.
For example, the 4am project~\cite{4am} gathers a collection of photos from around the world at the time of 4am, which has more than 7000 images from over 50 countries so far.
The ``Someone Once Told Me'' story~\cite{someone} collects images in which people hold up a card with a message that someone once told them.
In addition, shared photo album is an increasingly popular feature among photo sharing sites (e.g. Facebook) and mobile apps (e.g. Adobe GroupPix), which allows people participated in the same event to contribute to a shared album. 
All these collaborative applications focus on the sharing and storytelling part of the photography experience, but not the on-the-spot capturing experience.

PhotoCity is a game for reconstructing large scenes in 3D out of photos collected from a large number of users~\cite{Tuite:2011}.
It augments the capturing experience by showing the user the existing 3D models constructed from previous photos, as a visualization to guide the user to select the best viewpoint for taking a photo.
In this way, photos of different users can be combined together to form a 3D model.
However, this system is designed for {\em asynchronized} collaboration, meaning that photos are taken at different times.
In contrast, our system focuses on {\em synchronized} collaboration for applications that require all images to be taken at the same time. 

\section{Design Guidelines}

As mentioned earlier, the motivation of our  system is two folds.
On the one hand, we wish to provide a new photography functionality that enables capturing panoramas of highly dynamic scenes using regular mobile devices. 
On the other hand, we want to make this tool intuitive and fun to use, and to encourage interactions among people who are at the same location at the same time regardless of their current social relationship. 
To realize these two goals, we design our system with the following guidelines:

{\bf Encourage social interactions among users with a collaborative interface.} 
    As mentioned above, we propose to encourage more social interactions in the photography process by developing a collaborative capturing system, where every user contributes to only a portion of the final image. 
    
{\bf Provide a clear, easy-to-follow workflow.}   
    Panorama stitching is a fairly complicated process even for a single user.  Having multiple users to work collaboratively in a realtime capturing session requires a clear workflow where each step is intuitive and easy to follow. The types of actions and interactions in each step should be limited so users can quickly learn them on the spot, with limited training and practicing. 
    
{\bf Provide clear roles and tasks for different users.}    
  Given each user equal responsibility and capability in a single session may introduce conflict and confusion. It is desirable to have a single host user that plays the role of coordinating all other users and making sure each step is done correctly. Other users then play a supporting role and follow instructions given by the host . The two types of roles need to be provided with a separate set of user controls.     

{\bf Provide live preview and keep users in the loop.}
    When capturing panoramas using traditional tools, the users generally have little idea on how the final image will look like until the capturing is completed. 
    It would be especially frustrating if the system fails to generate a good result after intensive teamwork. 
    Thus, the system needs to constantly keep users in the loop, giving them immediate visual feedback after any camera movement, as well as a live preview of the final result. 

{\bf Limit the duration of a capturing session.}
   The longer a capturing session takes, the more likely that users may feel frustrated or tired and give up in the middle. 
   Thus, the system should be able to converge quickly, e.g. within 1 minute.

\section{User Experience}
\begin{figure*}[t]
  \centering
  \includegraphics[width=0.76\textwidth]{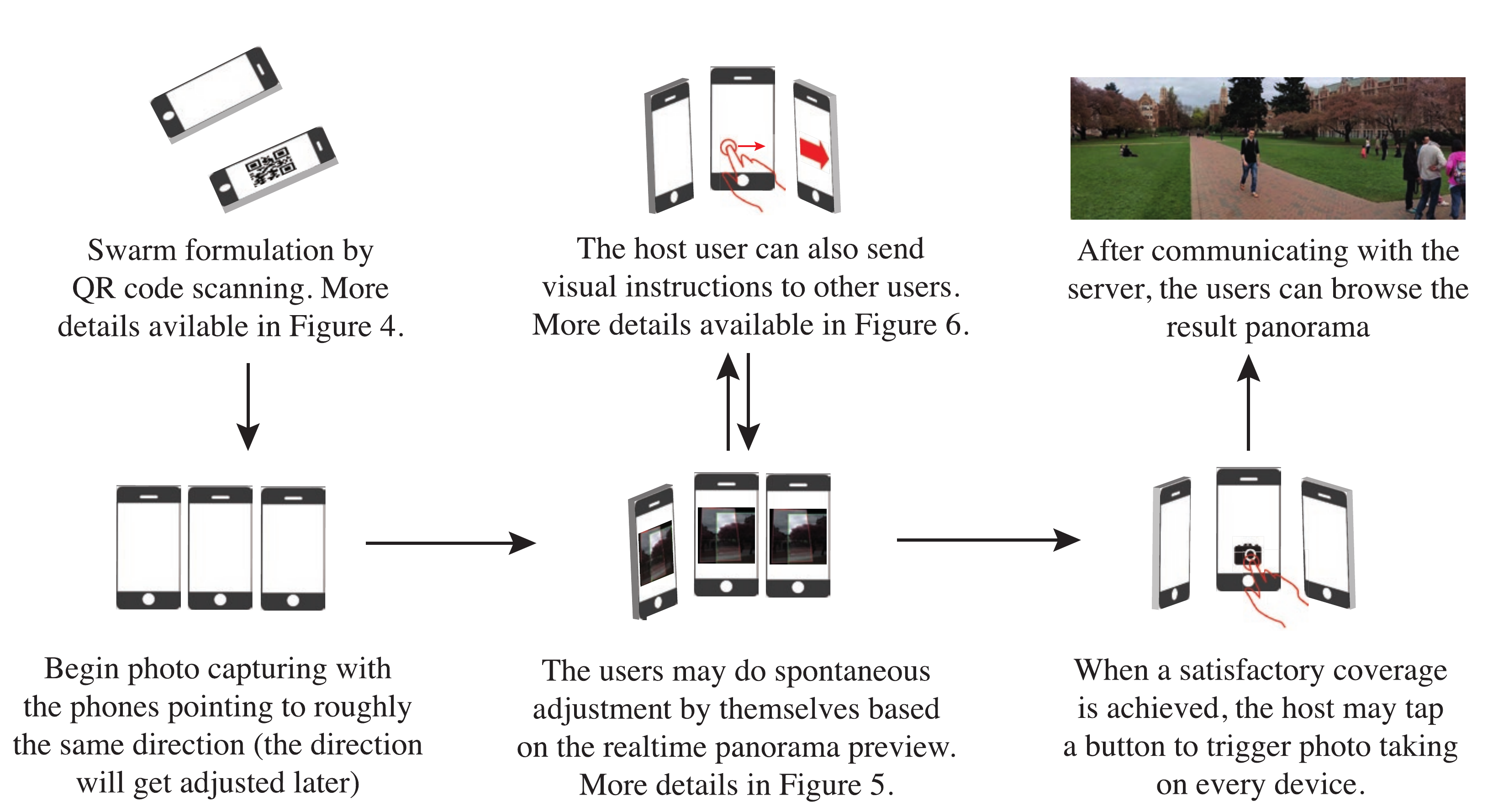}
  \caption{Flow chart of our PanoSwarm system.
  For clarity, only the critical components for each procedure are rendered on the phone screens in the figure.
 }
  \label{fig:userexp}
\end{figure*}

Following the design guidelines listed above,
we created PanoSwarm, a collaborative panorama capturing application
that lets users work together in a synchronized fashion for capturing and stitching a panorama.
Let us follow a group of users Alice, Bob and Carol as they use PanoSwarm to capture a panorama of an interesting and dynamic scene.

\subsection{Team Formation}
Alice, Bob and Carol all have PanoSwarm app installed on their iPhones.
Carol first opens the app on his iPhone, and selects the option of starting a new capturing session, which automatically makes him the \emph{host user} of the session.
A unique QR code then appears on his screen, allowing Alice and Bob to scan using their devices and join the session.
Alice scans the code first, and the same QR code automatically appears on her screen, so she can pass it to Bob. 

\subsection{Guided View Finding}
After all three users join the session, they make a selection on the UI to enter the capturing mode.
Initially, they point their cameras to roughly the same direction, so the system can determine their relative camera positions based on the overlapped portions of the images, and then they begin to adjust the camera directions with the help of the interface.
On each user's screen, a preview panorama automatically appears, with colored bounding boxes showing the contribution from each camera.
Being the host, Carol then guides Alice and Bob to adjust their cameras to increase the scene coverage of the panorama. 
He selects Alice's camera on his screen and swipes to the left. 
On Alice's screen, she immediately sees a red arrow pointing to the left, indicating that she is instructed to turn her camera that way.
She then gradually moves her camera towards the left, and sees that the panorama is updated in realtime according to here camera motion. 
The red arrow only exists for a short period of time and then disappears.
Carol monitors Alice's camera motion on his own screen, and he feels the movement is not enough.
So he keeps sending the swipe gesture to Alice to instruct her to keep moving, until her camera is turned into the desired direction. 

Similarly, Carol selects Bob's camera and use the swipe gesture to instruct him to turn his camera to the right, which Bob follows. 
However, Bob moves his camera too far and his image can no longer be stitched with other ones. This is reflected on the panorama preview, and Bob notices it and moves his camera back. 
Finally, Carol directly talks to Alice and Bob to move their cameras up a little bit, to capture more on the building and less on the ground.  

\subsection{Capturing}
When Carol feels the current preview panorama is good to capture, he clicks the button to trigger the capture event. 
Alice and Bob simultaneously see a countdown on their own screen, and they keep their cameras still before the countdown reaches zero. When it happens, all cameras take pictures at the same time. 
After a short delay, they then stream the high quality panorama from the server for browsing. 

The overall user experience chart is also provided in \Fig{userexp} for better illustration.

\section{System Description}

\begin{figure}[t]
    \centering
    \includegraphics[width=0.5\textwidth]{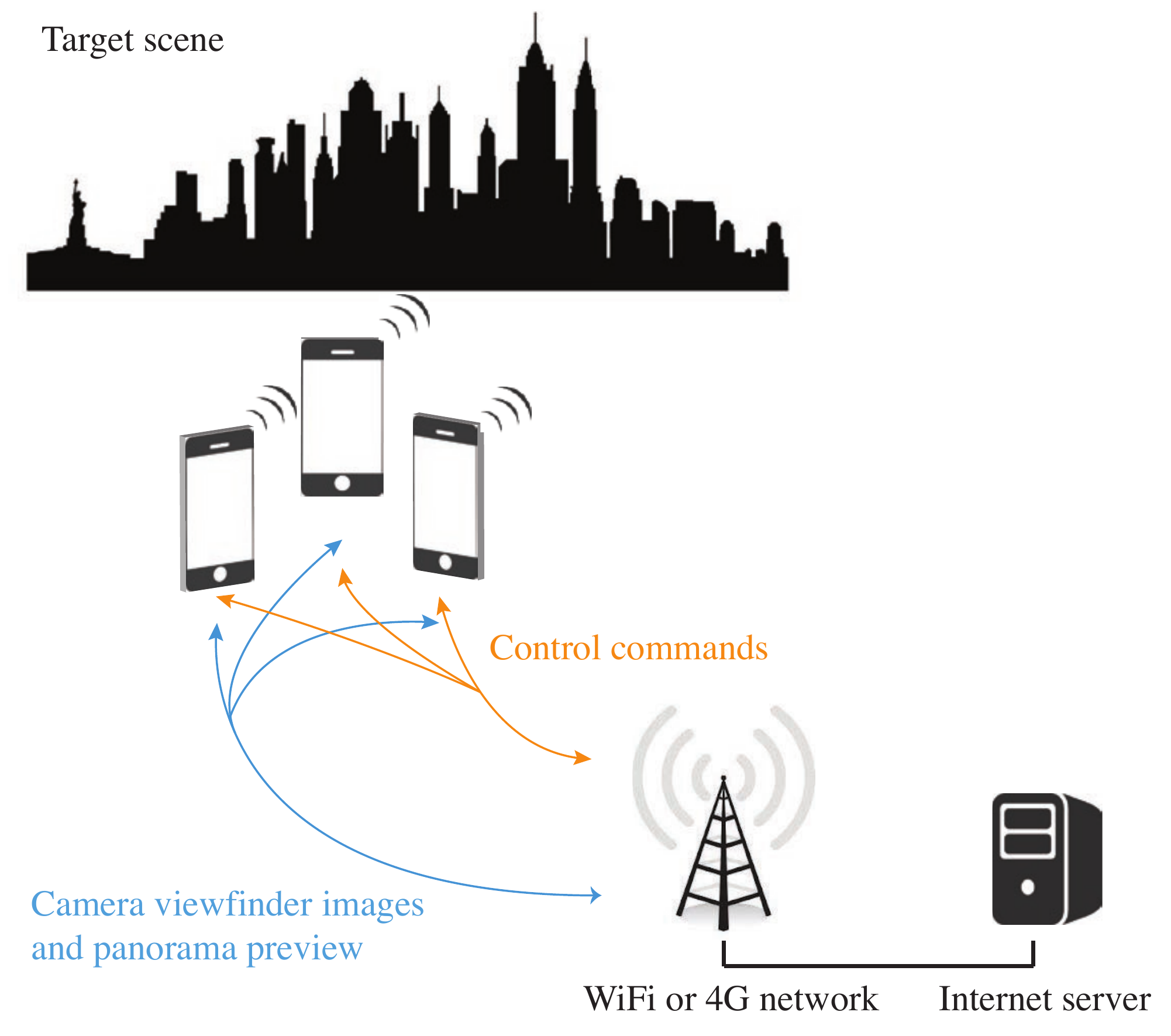}
    \caption{Technical framework of the proposed system.}
    \label{fig:framework}
    \vspace{-0.2cm}
\end{figure}

The PanoSwarm system consists of two components: (1) an Internet server that communicates with each participating devices; and (2) a mobile app (currently implemented on the iOS platform) that runs on each device. The server and the app communicate using standard HTTP protocols.  Given the amount of data being transmitted in realtime, the system requires WiFi or 4G network to be available to function smoothly,  which is usually not a problem for urban scenes. the system framework is shown in Figure~\ref{fig:framework}.

\subsection{Dynamic Team Formulation}

\begin{figure}[t]
    \centering
    \includegraphics[width=0.38\textwidth]{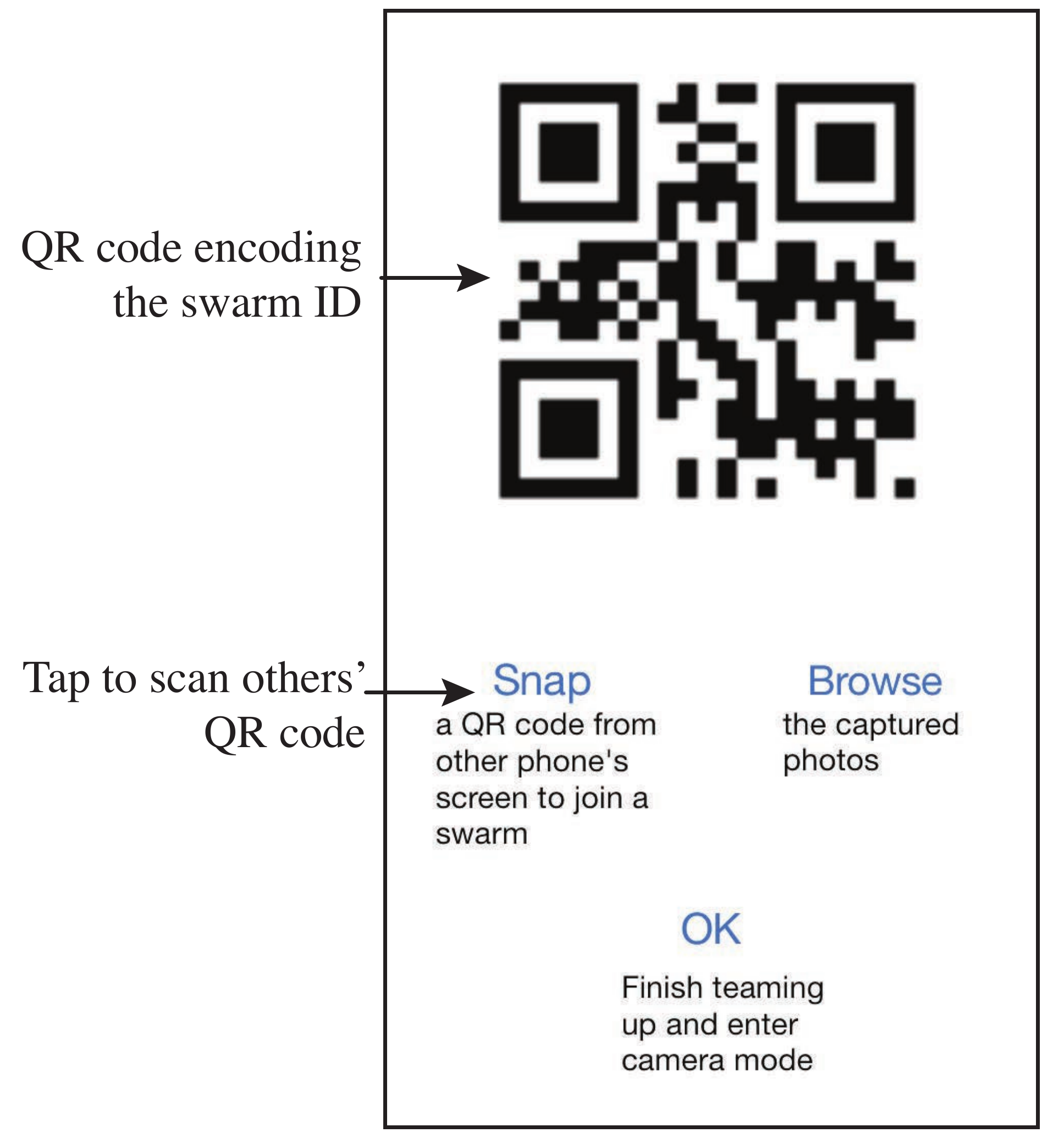}
    \caption{The main interface for easy team formulation.
    User can join other people's team by simply tapping the ``Snap'' button and scan others' QR code.}
    \label{fig:qrcode}
    \vspace{-0.45cm}
\end{figure}

Our system allows users that are physically at the location to dynamically and temporally form a team for a capturing task. 
This flexible setup allows users who even do not know each other before to work together as a team. 
Specifically, when the app launches, it registers itself with the server and obtains a group ID, which
is encoded in a QR code and being displayed on the main interface, as shown in \Fig{qrcode}.
Note that every user forms a minimal group consisting of only one user when the app is launched.

A user A can then scan another user B's QR code to join the group that B is currently in. 
After scanning, device A is assigned with the same group ID as B, and A's QR code on the main interface is replaced by the QR code of B.
After A and B are in the group, they can both add news users into the session. In this way a large group can be quickly formed. 

On the server side, the server program keeps track of which devices (represented as a unique combination of IP address and port)  are associated with the same group ID, and among them which one is the host. When the host taps the ``OK'' button on the UI to end team formation and enter the actual capturing process, a unique section ID is assigned to this group. 

\subsection{Live Panorama Preview}

During the capturing process, each participating device sends a downsized version (maximum dimension is $300$ in our implementation) of the current viewfinder image to the sever at the rate of 3 frames/second. The server collects images from all devices and generate a coarse preview panorama in realtime on-the-fly, which is then streamed back to each device. We use low resolution images in realtime communication in order to prevent network congestion and server processing burden, so as to maintain a smooth user experience.

The preview panorama serves for two main purposes. First, it gives each user a direct visualization of how his/her own image contributes to the final panorama, and in which direction he/she should move the camera to increase the scene coverage. Without such preview, it would be really hard for users to mentally reconstruct the relative camera positions, even they could see the viewfinder images of other devices.  

The online preview also turns panorama capturing into a WYSIWYG experience. 
In the traditional panorama capturing workflow, the user is required to finish capturing all the images at first, then invoke an algorithm to stitch the panorama offline. However, given panorama stitching is not a trivial task and involves using advanced computer vision techniques, it may fail at times, and the user either has to give up or re-capture the whole series of images. 
The lack of instant feedback makes this task to be unpredictable.  
With the live preview provided by our system, the user can instantly see how his/her camera motion affects the final result, and has the opportunity to adjust the camera to avoid/correct any errors or artifacts in the final result, before capturing the actual images. 
It thus significantly increases the success rate of the system.  
This is particularly important for collaborative teamwork, as it takes a significant amount of effort for all participating users to accomplish a collaborative session, thus a bad result is particularly discouraging. 

Computing panorama in realtime is technically challenging, even for downsized images. To maintain computational efficiency, we implemented a simplified version of the panorama stitching algorithm on the server, which only uses SIFT feature matching to estimate affine transforms for alignment. Other advanced operations, such as exposure correction, lens distortion correction, and seamless blending are not performed in the preview stage, but only in the final rendering stage. 
Such stitching is performed whenever a new viewfinder image is received by the server.
We implemented this method using C++ on a server machine with a Core i7 3.2GHz CPU, which achieves $20$ fps panorama updating for one PanoSwarm session with four clients.


\subsection{Guided Camera Adjustment}

As mentioned earlier, the users start a capturing session by pointing their cameras to roughly the same object. Then, based on the guidance of the live panorama preview, the users adjust each individual cameras to increase the scene coverage. The adjustments can be made either spontaneously by individual users, or under the guidance of the host user, as we will discuss in this session. 

\begin{figure}[t]
    \centering
    \includegraphics[width=0.32\textwidth]{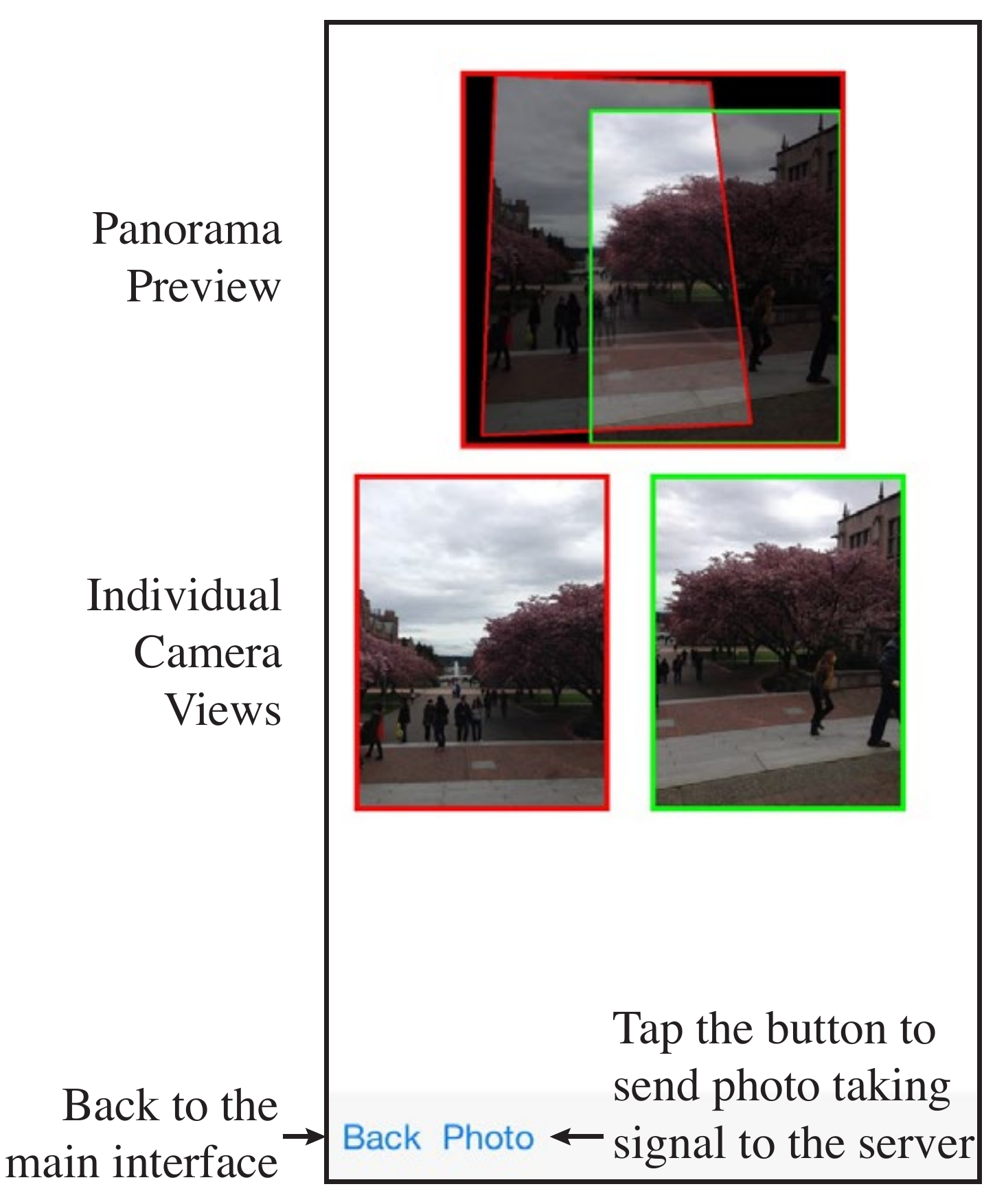}
    \caption{The user interface for the spontaneous adjustment.}
    \label{fig:panopreview}
\end{figure}

\textbf{Spontaneous Adjustment.}
In addition to the live panorama preview, our system also provides additional visualization to help users make spontaneous camera adjustment. 
As shown in \Fig{panopreview}, overlaid on the live panorama preview in the upper half of the interface, each user's camera is visualized within a colored bounding box. 
At the lower half of the interface, each user's viewfinder image is displayed in boxes in corresponding colors.
Each user is assigned with a unique color, and a user can quickly identify his/her own color by looking at the border of the panorama preview in the upper half, thus to find the corresponding scene coverage in the panorama. 
It is also easy for the host user to find out the scene coverage of any devices by reading the colors of the bounding boxes. 

Once a specific user identifies his/her color, he/she can then move the camera gradually away from the common initial position to increase the scene coverage.  
Since the preview is updated in realtime, the user instantly sees the impact of the movement to the final result, as well as the relative camera positions of all the devices. 
The goal of each user is to maximize the coverage of the panorama preview, while maintaining a reasonable amount of scene overlap (typically 1/5 of the size of the image) with adjacent cameras, which is required by the stitching algorithm.


\textbf{Instruction-based Adjustment.}
One problem for spontaneous adjustment is that different users may start conflicting or duplicating adjustments at the same time. 
For instance, after initialization, two users may start moving their cameras towards the same direction.
After they realize that someone else is doing the same adjustment, they may turn their cameras back at the same time again. 
Thus, the lack of coordination may make the process hard to converge. 

\begin{figure}[t]
  \subfigure[The host user swipes up-left on the camera view of the other user, suggesting her to turn the camera up-left.]{
    \includegraphics[width=0.22\textwidth, trim=0 0 5 0]{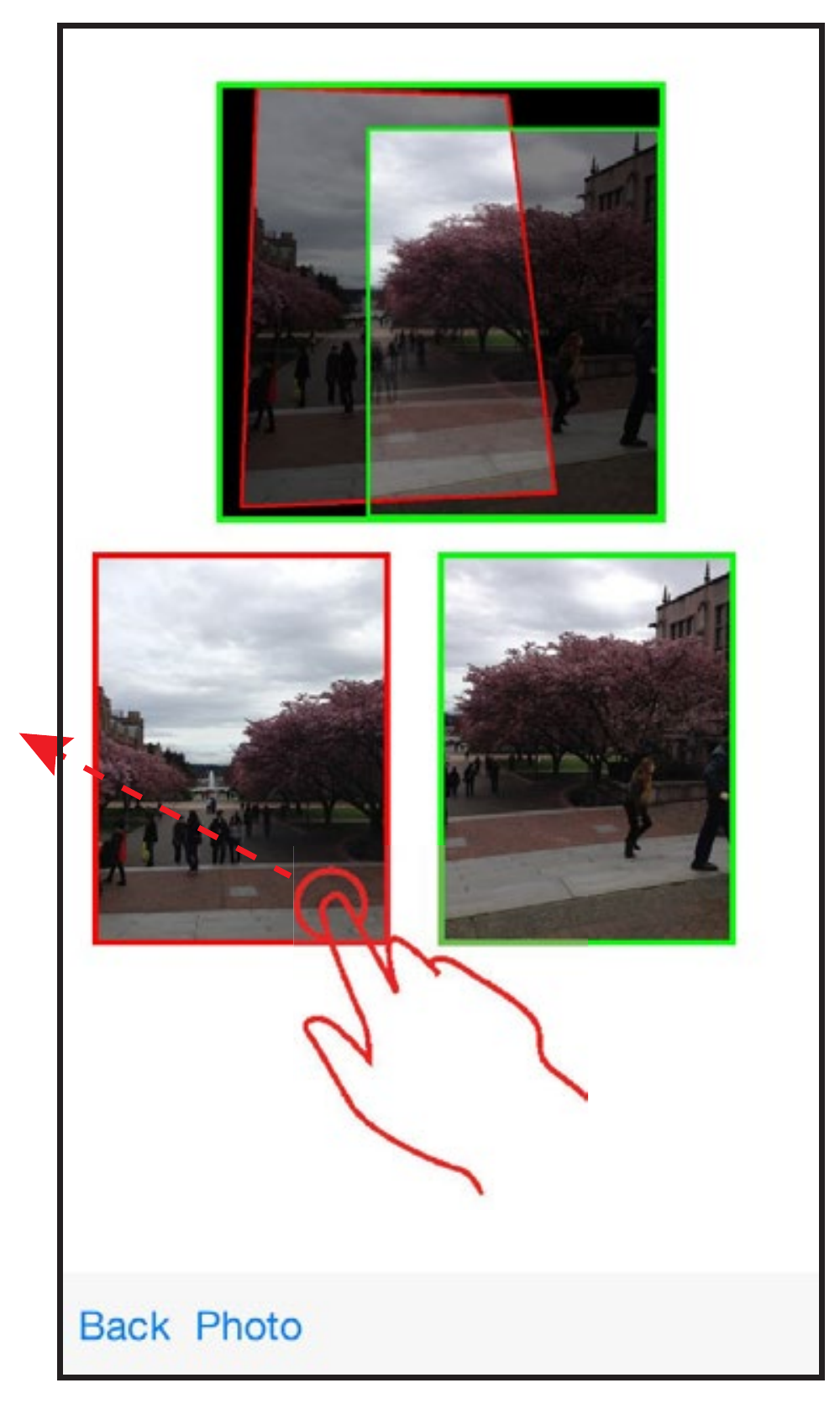}
    \label{fig:master:master}
  }
  \hfill
  \subfigure[A red arrow is rendered on the user's screen to prompt the host user's suggestion.]{
    \includegraphics[width=0.215\textwidth, trim=0 0 5 0]{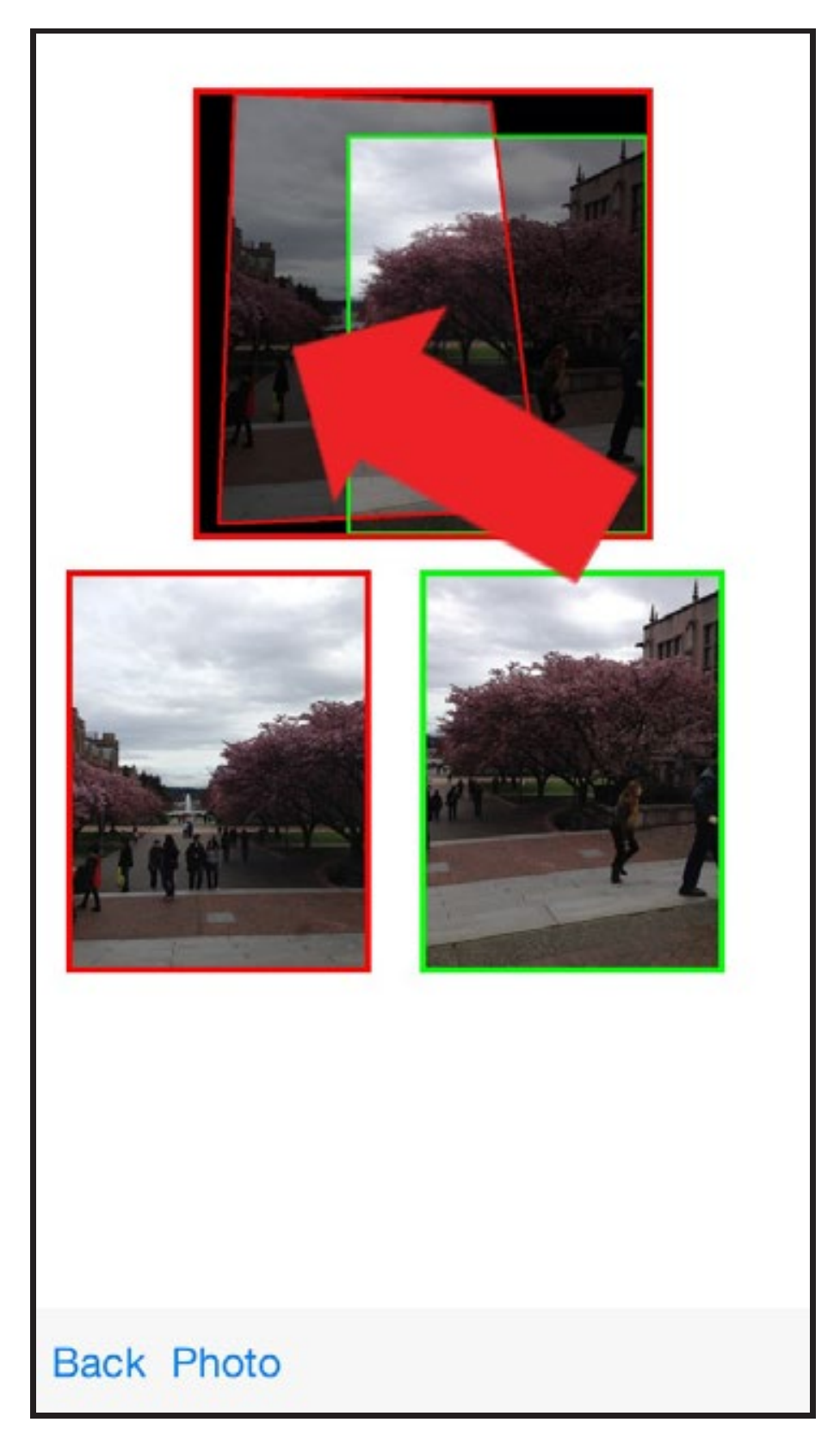}
    \label{fig:master:master2}
  }
  \caption{The interface showing the host led adjustment process.}
  \label{fig:master}
  \vspace{-0.3cm}
\end{figure}

To avoid this problem, our system allows the host user to directly give instructions to other users to guide the overall camera adjustment process. 
Specifically, when the host user finds a user A is contributing too little to the panorama (e.g. because of too much overlap), he/she can easily identify user A's camera in the second row in the interface, based on the color of the bounding box.
By simply touching A's camera view and swiping along the direction that the host wishes A to move, a command containing the target user (i.e. A) and the swiping direction will be sent to the server, and then forwarded to A's device.
Once the instruction is received on A's device, a red arrow will be rendered on A's screen, prompting the user to turn the camera to the suggested direction.
The red arrow has a relatively short lifespan and will disappear shortly. If the host user wishes a large movement of A's camera, he/she can swipe multiple times to encourage A to continue the movement until desired camera orientation is reached. 
This process is illustrated in \Fig{master}, with \ref{fig:master:master} indicating the host's view, and \ref{fig:master:master2} showing user A's screen.

\subsection{Photo Capturing}

Once all the cameras have reached their desired scene coverage, the host user can trigger the photo capturing event by tapping a button on the user interface.
As discussed earlier, allowing all devices to capture simultaneously is essential for taking motion-artifact-free panoramas in highly dynamic scenes.
Our approach for achieving this is to send a signal to the server, and the server will forward this signal to all the devices in the same session.
Once a device receives the signal, the app automatically triggers the camera to take a picture, and then upload it to the server for final panorama stitching.

\begin{figure}[t]
  \centering
  \includegraphics[width=0.5\textwidth]{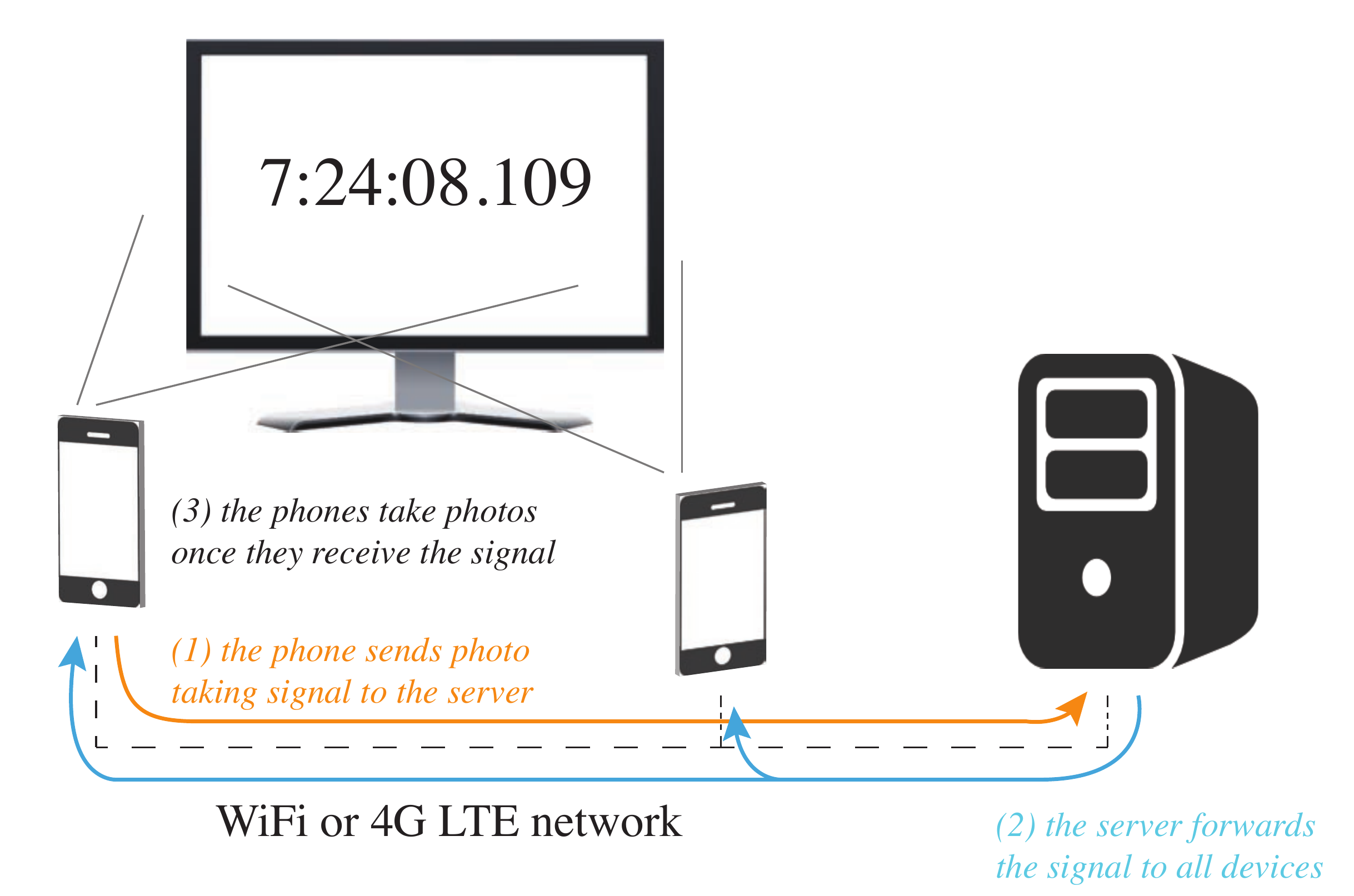}  
  \caption{The experiment to verify the effectiveness of the proposed approach of simultaneous photo capturing.}
  \label{fig:simulexp}
\end{figure}

To verify whether this approach can produce accurate synchronization, 
We first test its feasibility in a controlled environment. 
As shown in \Fig{simulexp}, we display a millisecond level clock on a computer screen, and set up two smartphones that both point to the clock.
We implement the signal forwarding system mentioned above, and let one of the phones send out the photo taking signal to the server. Both phones will immediately take a photo of the clock once they receive the signal form the server, including the phone that sent out the signal initially.
Using this design, variance of the signal arrival delay between the server and the clients can be directly measured, by comparing the difference of the time shown in the photos taken by individual cameras.

We conducted this experiment using WiFi connection as well as 4G LTE networks, and observed time differences on different devices to be well below $100$ms.
In WiFi networks, the difference is generally smaller than $40$ms. 
This small difference is usually negligible in real panorama capturing tasks, and would not introduce noticeable artifacts.   

\section{Preliminary Evaluation}

We implement the proposed system as an iPhone app, and provide a video to demonstrate the system in action, available at \url{http://youtu.be/PwQ6k_ZEQSs}.
We designed our evaluation to answer three questions: 
can PanoSwarm help user capture good panoramas of challenging scenes?
Does the proposed interface intuitive to learn and use?
Furthermore, can PanoSwarm effectively encourage social interactions among people?

\subsection{Methodology}

We compared two interfaces: the built-in panorama capturing procedure in iOS, and the proposed PanoSwarm system.

{\bf Participants.} We conducted two user study sessions. For the first session, we recruited four users (A1-A4, all male, user group A) from an open
e-mail list at a private university. For the second one, we recruited another three users (B1-B3, one female, user group B) at a large public university. 
All participants self-reported that they extensively used their mobile devices for their photography needs, but had diverse experiences with panorama capturing.
Three users were enthusiasts that took panoramas at least once a week (A1, A2 and B1). Three users were casual users that tried a couple of times a year (A3, A4 and B2). User B3 had never tried capturing a panorama before.   
In addition, the age of the participants ranged from 21 to 32.

{\bf Training.} At the beginning of each user study session, 
we trained our participants on how to use our system by going through all the steps once in a capturing session.
One of the authors played the role of the host user and explained every step verbally and answered questions, until the subjects had no more questions and can take proper actions in each step. 
We also explained to user B3 the basic knowledge of panorama stitching, given that he had never done it before. 

{\bf Tasks.} Immediately following the training session, we asked participants to complete at least three capturing sessions of different scenes.
When arrived at each scene, before using our system, we asked for one volunteer to take a panorama using the built-in panorama stitching tool in iOS. 
For each capturing session, we tried to intervene as little as possible, mostly just helped fix some occasional technical issues such as network and device failure.  
We let the users themselves to decide who should be the host user and which part of the scene to photograph. 
We also did not give the users specific goals to achieve.

{\bf Debrief.} We debriefed the participants after they completed their tasks.
We asked them for their overall impressions, how PanoSwarm compares to iOS panorama stitching tool, and their overall feedback.
In addition, we also asked the participants to complete a short questionnaire.

\subsection{Results and Findings}

In our study user group A finished three capturing sessions and generated three panoramas, and user group B produced six panoramas in six sessions. 
Every capturing session resulted in a reasonable panorama without dramatic failure such as failed alignment or producing large holes in the panorama, indicating the direct and live panorama preview in the capturing process can effectively help users prevent disasters. 

\begin{figure*}[t]
  \centering
  \includegraphics[width=0.8\textwidth]{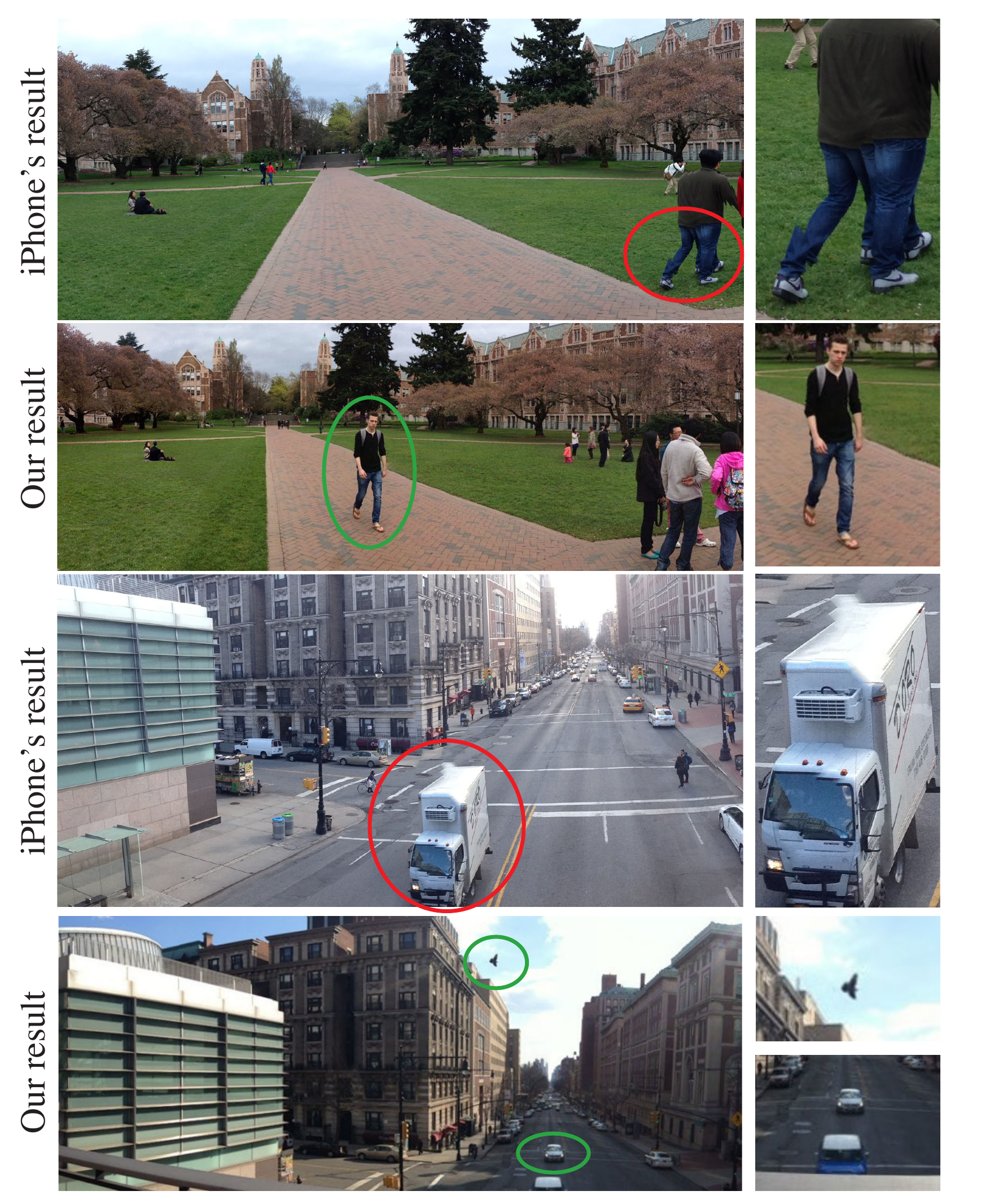}
  \caption{Comparisons of the panoramas produced by iPhone (top) and our proposed system (bottom) on dynamic scenes. 
  From top to the bottom, each row shows iPhone's result in scene A, our result in scene A, iPhone' result in scene B, and our result in scene B.
  Note the ghosting artifacts (highlighted in red) in iPhone's results, and the fast-moving, artifact-free objects (highlighted in green) in our results, which are also highlighted in the right column.}
  \label{fig:panocom}
\end{figure*}

In \Fig{panocom} we show a few panoramas captured by the subjects and compare them with the ones captured by the built-in iOS tool. It suggests that our system can faithfully capture all dynamic objects in the scene, thanks to the simultaneous shutter release on all the devices. In contrast, the iOS tool had difficulties to deal with moving objects and produced obvious ghosting artifacts. It also shows that the differences in exposure time among the devices are small enough to produce high quality results. 

\begin{table}[h]
\centering
\caption{The duration of each capturing sessions of the two user groups. }
\label{tab:time}
\begin{tabular*}{0.4\textwidth}{c|c|c}
\hline
  & \# of users & Time (in seconds)   \\
\hline
  Group A & 4 & 67, 54, 57 \\
  Group B & 3 & 38, 24, 26, 23, 24, 28 \\
\hline
\end{tabular*}
\end{table}

Table~\ref{tab:time} shows the duration of each capturing session.
Some interesting observations can be made from this data:

\begin{enumerate}
\item For both user groups, the first session took noticeably more time than the following sessions, possibly indicating that the users need more time to get familiar with the interface and workflow when using the app for the first time. 
\item After the first session, the duration of the capturing sessions quickly came down and became relatively stable, suggesting that the users can learn how to use the app fairly quickly.
\item The average session duration for group A is almost twice as long as that of group B. While the two groups consist of different users, we hypothesize that this is largely due to the different sizes of the groups. Group A has one more user, thus it may have significantly increased the required user effort for collaboration. 
\end{enumerate}

\subsubsection{Ease of Use}
We also asked about the subjects' opinions about the ease of use of the interface, as well as how they like it compared to iPhone's panorama, in the questionnaire.
All users agreed that the system is easy to learn, with better capability than iPhone's default tool, except B3, who had not used any panorama tool before. User B2 commented that ``{\em iPhone does a great job for static scene panoramas. however, it is not very good in two things:
1. Lack of ability for dynamic scene, e.g. ghost shadow.
2. You need to hold steady while moving the phone.
PanoSwarm does a good job in both of these.}'' User A4 commented that ``{\em from the early prototype I have seen, I think this app fills a totally different need than the native iphone panorama. When messing with the native app you notice that people in frames sometimes get cut off because they are moving- we didn't see anything like that with this app. There is of course the overhead that you need more people with PanoSwarm, but like I mentioned earlier I think this is a cool feature}''. 

Since our interface provides multiple dynamic views, such as the panorama preview and the live viewfinder images from all devices, we asked the subjects which view is the most important and intuitive to them. The users' answers were highly consistent: they all agreed that {\bf the live panorama preview is almost the only view that they looked at during the capturing sessions}, and the live viewfinder images at the bottom of the interface were not particularly useful. This suggests that for future improvement, we could potentially remove all viewfinder images and focus on improving the live panorama view.

\subsubsection{Social Interactions}
In the evaluation, we noticed that although we did not explicitly require the users to make interactions when using our app, this process naturally generates social activities.
{\bf We observe that people naturally started to talk to each other in the process as the primary communication channel, while the host user also used the app to send out visual instructions}.
People talked primarily in the early stage of a capturing session, to figure out which area is his or her responsibility. After that the users mainly used the panorama preview to adjust their directions, with occasional instructions from the host.

{\bf When asked whether they would like to use the app in the future with their friends, all users commented positively}. User B3 never had any panorama capturing experience, but he commented that ``{\em maybe not on daily basis. It looks time-consuming for me to take photos like that. But in some situations I'd like to use it to keep memories, e.g., in an event with a lot of friends.}''. 

\section{Discussion}

While we are glad that the preliminary study shows that users found PanoSwarm effective and
useful, there are many ways we can improve it in future work. There are also many interesting questions left unanswered that we would like to address in future studies. 

\subsection{The Size and the Skill Level of the Team}

Our preliminary study only involved teams of $3$ and $4$ users. We observed that in general the $3$-user team had shorter capturing sessions than the $4$-user team ($27$ seconds v.s. $59$ seconds in average), given that there was one less camera to adjust.  
Increasing the number of users will increase the computational load of the server, which may make the server less responsive. 
However, more users in a session may make the task more challenging and potentially generate more rewarding panoramas with larger scene coverage. 
The questions that is interesting to us is, ignoring the computational issues, is there a sweet spot for the number of users in a single session to maximize the overall experience? 
Our hypothesis is that the answer to this question highly depends on the skill levels of individual users in the session. If all users are experts, then perhaps a larger team if preferred. On the contrary, if all users are novices,  then having smaller teams would reduce the chance of failure. It is however unclear what happens if the users have mixed skill levels. 
We plan to conduct more studies in the future to explore along this direction. 

\subsection{Can Strangers Work Together?}

The two user study teams in our preliminary study were formed by people that already know each other before.
Thus, there is less a barrier among them to work together to accomplish a task.
However, it is unclear whether total strangers are willing to team up, and furthermore, accomplish the task effectively.
We included the question ``do you wish to use PanoSwarm with strangers in the future'' in our questionnaires, and the turned-around opinions are mixed.
Some of the users show interests, for example, user A4 commented that ``{\em of course, I think this is an amazing experience to share with someone I do not know. Great way to break the ice/ meet people who share similar interests (b/c you guys are taking pics of the same stuff)}''.
But more users are inclined to only use it with friends.
An especially interesting view from user B3, who never used panorama stitching before, is that he prefers to not using the app with strangers, but is willing to show off to other people as a conversation starter by using it with his friends. 
As one of the main motivation of the project is to encourage social interactions among people who otherwise would not interact with each other, we plan to conduct future studies to evaluate its effectiveness on doing so. 

\subsection{Other Collaboration Modes}

Our current app has a fixed setup where one host user controls the overall capturing process. 
When asked what can be further improved, one subject commented that ``{\em I'm not too keen on the idea of a single host controlling where others should move. I think there should be a way to make it more fun for all the users.}''
This inspires us to develop an alternative collaboration mode, where each user has equal responsibilities and power in the capturing process. For example, instead of sending instructions to other users for camera adjustment, each user can send all other users a signal indicating where he/she plans to move the camera, and make sure it does not conflict with other users' intentions that have already been broadcasted.  
This setup does require more calibrated collaboration, but may be more fun to advanced use and can provide  a stronger sense of accomplishment.
We plan to implement this mode and further study it. 

\section{Conclusion}

We presented PanoSwarm, the first collaborative photography tool that helps multiple users work together to capture high quality panoramas of highly dynamic scenes, which cannot be done with traditional single camera panorama capturing workflow.
Our system employs a client-server framework.
The server collects all viewfinder images and stitch them on-the-fly for live panorama preview.
The client interface shows the preview to the user and allows the host user to send visual instructions with gestures to other users for coordinated camera adjustment.
Our preliminary user study suggests that the proposed system can provide a unique, enjoyable experience for users in addition to capturing high quality panoramas. 

\balance 

\bibliographystyle{acm-sigchi}
\bibliography{main}
\end{document}